# Evolution of the phase singularities in edge-diffracted optical-vortex beams

Aleksandr Bekshaev, Lidiya Mikhaylovskaya  
Research Institute of Physics  
I.I. Mechnikov National University  
Odessa, Ukraine  
bekshaev@onu.edu.ua

Aleksey Chernykh, Anna Khoroshun  
Physical Department  
East Ukrainian National University  
Severodonetsk, Ukraine  
an_khor@i.ua

*Abstract*—We study, both theoretically and by experiment, migration of the amplitude zeros within a fixed cross section of the edge-diffracted optical-vortex beam, when the screen edge performs permanent translation in the transverse plane from the beam periphery towards the axis. Generally, the amplitude zeros (optical-vortex cores) describe spiral-like trajectories. When the screen edge advances uniformly, the motion of the amplitude zeros is not smooth and sometimes shows anomalously high rates, which make an impression of instantaneous "jumps" from one position to another. We analyze the nature, conditions and mechanism of these jumps and show that they are associated with the "birth–annihilation" topological reactions involving the optical vortex dipoles.

## I. Introduction

For a long time, edge diffraction of optical vortices (OV) has been an actual topic enabling the spectacular manifestation of the special OV properties associated with the transverse energy circulation (TEC) [1–9]. One of the most impressive evidences of the TEC is the recently revealed spiral-like motion of the OV cores within the diffracted beam cross section, that occurs when the screen edge performs a monotonous translation in the transverse direction towards or away from the beam axis [10,11] (see Figs. 1, 2).

Besides the general spiral-like evolution, trajectories of the OV cores in the observation plane show other intricate peculiarities (see in Fig. 2a–c the behavior of three secondary OVs formed due to diffraction of the third-order circular Kummer beam [12] generated by the "fork" hologram from the incident Gaussian beam of radius *b*). The most interesting is the anomalously rapid evolution in some trajectory segments, for example, those marked by the cyan stars. In Figs. 2a–c, the total distances between the stars correspond to decrements of *a* from 3.76*b* to 3.73*b* (blue curve), from 3.15*b* to 3.05*b* (black curve) and from 2.36*b* to 2.34*b* (red curve), correspondingly. Importantly, in these intervals only one of the secondary OVs moves rapidly while positions of the two others are practically unchanged. Likewise, in other segments of trajectories, comparable changes of *a* cause 1 – 2 orders of magnitude smaller OV displacements. In this report, we investigate the nature and mechanism of this effect.

## II. Analytical model

In the coarse approximation that is, however, satisfactory under conditions of small beam perturbation (which is realized when the screen edge is separated by several *b* from the incident beam axis, see Fig. 1), the diffracted beam can be considered as a superposition of the unperturbed incident beam with the near-axis amplitude distribution

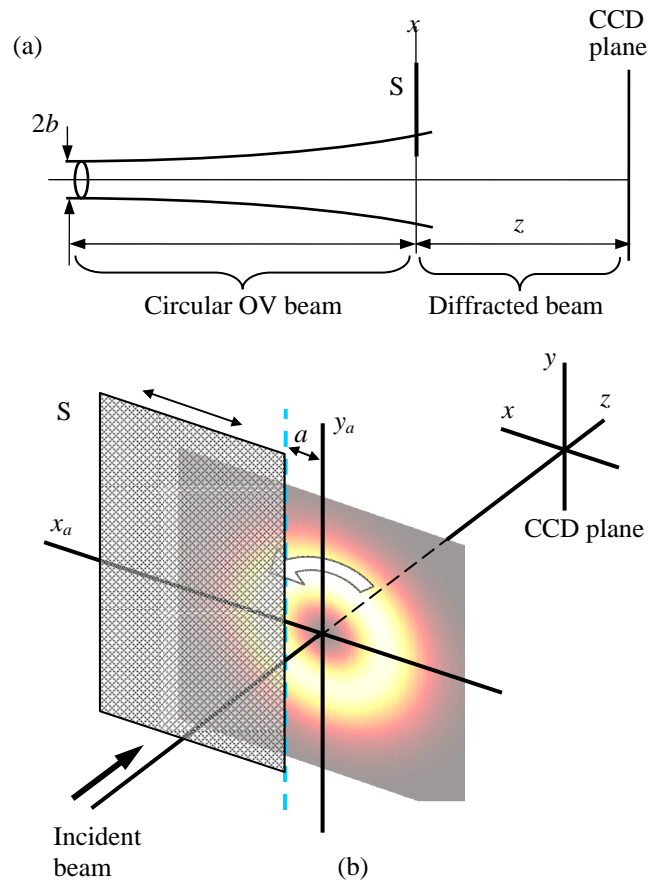

Fig. 1. (a) Scheme of the OV beam edge diffraction and (b) magnified view of the beam screening and the involved coordinate frames. S is the diffraction obstacle (opaque screen with the edge parallel to axis *y* and whose position along axis *x* is adjustable), the diffraction pattern is registered in the observation plane by means of the CCD camera.

$$E_{\text{inc}} = B_0 \left(\frac{r}{b}\right)^{|m|} \exp(im\phi)\exp(ikz) \qquad (1)$$

($m$ is the incident OV topological charge, $k$ is the radiation wavenumber, $r$ and $\phi$ are the polar coordinates in the CCD plane, $B_0$ is a certain complex constant) and the edge wave "emitted" by the screen edge. Near the origin of the observation plane its amplitude approximately amounts to

$$E_{\text{edge}} = D_0(a)\exp\left[ik\left(z + \frac{a^2}{2z} - x\frac{a}{z}\right)\right]$$

$$= D_0(a)\exp\left[ik\left(z + \frac{a^2}{2z} - r\frac{a}{z}\cos\phi\right)\right] \qquad (2)$$

with complex coefficient $D_0(a)$ decreasing with growth of $|a|$. Eq. (2) differs from the expression used in [11] by the $x$-proportional term responsible for the wavefront inclination in the ($xz$) plane (see Fig. 1). Positions of the OV cores are determined by the condition $E_{\text{edge}} + E_{\text{inc}} = 0$, which entails

$$\frac{r}{b} = \left[\frac{|D_0(a)|}{|B_0|}\right]^{1/|m|}, \qquad (3)$$

$$\phi + \frac{kra}{mz}\cos\phi = C_{N+1} + k\frac{a^2}{2mz} \qquad (4)$$

where the coordinate-independent term $C_{N+1}$ possesses its own value for each secondary OV numbered by $N = 0, 1, \ldots |m|-1$,

$$C_{N+1} = \frac{1}{m}\left[\arg D_0(a) - \arg B_0 + (2N-1)\pi\right]. \qquad (5)$$

The azimuthal coordinate of the OV core is determined by Eq. (4) which, in contrast to its counterpart Eq. (13) of [11] is now transcendent. Its qualitative analysis is illustrated by Fig. 3a. The dependence of the left-hand side expression on $\phi$ is imaged by the blue curve, points of the vertical axis express the values of the right-hand side, the equation solution $\phi(a)$ is obtained as intersection of the blue curve and the horizontal line corresponding to a given $a$. In normal situation for a given $N$, there is only one intersection point (see, e.g., points $\phi_1$ and $\phi_4$ in Fig. 3a). When applied to the case of $m < 0$ presented in Fig. 2, with $a$ decreasing monotonically, the horizontal line moves upward, and the corresponding $\phi(a) = \phi_1$ also changes monotonically and continuously.

However, due to the trigonometric term in Eq. (4), the blue curve can be non-monotonic, and at certain values of $a$, the horizontal line reaches the region where the left-hand side decreases (between the red dashed lines in Fig. 3a). Obviously, in this region $\phi(a)$ can change very rapidly; besides, there appear additional intersections (see the green line) that witness for emergence of additional OVs.

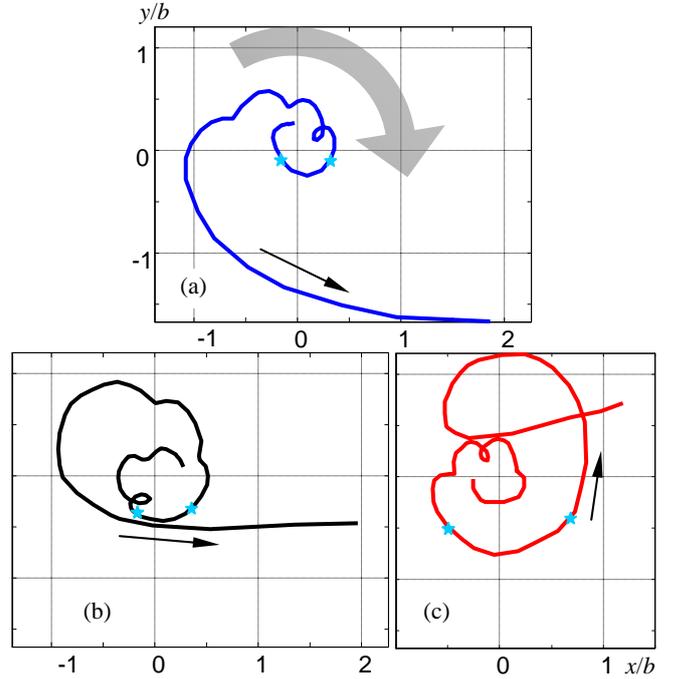

Fig. 2. Trajectories the OV cores in the diffracted beam cross section $z = 30$ cm behind the screen when the screen edge moves towards the negative $x$ (see Fig. 1b), for the incident Kummer beam with topological charge $m = -3$. The horizontal and vertical coordinates are in units of $b$; the large grey arrow shows the energy circulation in the incident beam, small arrows show the direction of the OV motion

### III. NUMERICAL ANALYSIS

Figs. 3b–d present the numerical example explaining the behavior of the "red" secondary OV whose trajectory is shown in Fig. 2c, between the points denoted by the cyan stars. The OV positions can be recognized as points where several equiphase contours converge together and the phase surface "cuts" (congestions of nearly coinciding lines of different colors) end; they are marked by curve arrows showing the local TEC, colored in compliance with the trajectory colors in Fig. 2. The situation when, due to decrease of $a$, the right-hand side of Eq. (4) approaches the non-monotonic region (which corresponds to the left cyan star in Fig. 2c and to the point $\phi_2$ in Fig. 3a), there are three secondary OVs distinctly seen in Fig. 3b. At this moment, a small advance of the screen towards the axis almost does not affect the OV positions but induces a topological event: in the area indicated by black circle in Fig. 3b, the cut is torn and the dipole of oppositely charged OVs emerges (see. Fig. 3c). With further decrease of $a$, one of the new-born OVs, charged oppositely to all the other OVs (green curve arrow), rapidly moves along the "inter-star" segment of the trajectory of Fig. 2c, against the "normal" spiral OV motion. Then it meets the "red" OV A and annihilates with it, whereas the second member of the dipole pair, B, still exists as a "continuation" of the OV A (Fig. 3d).

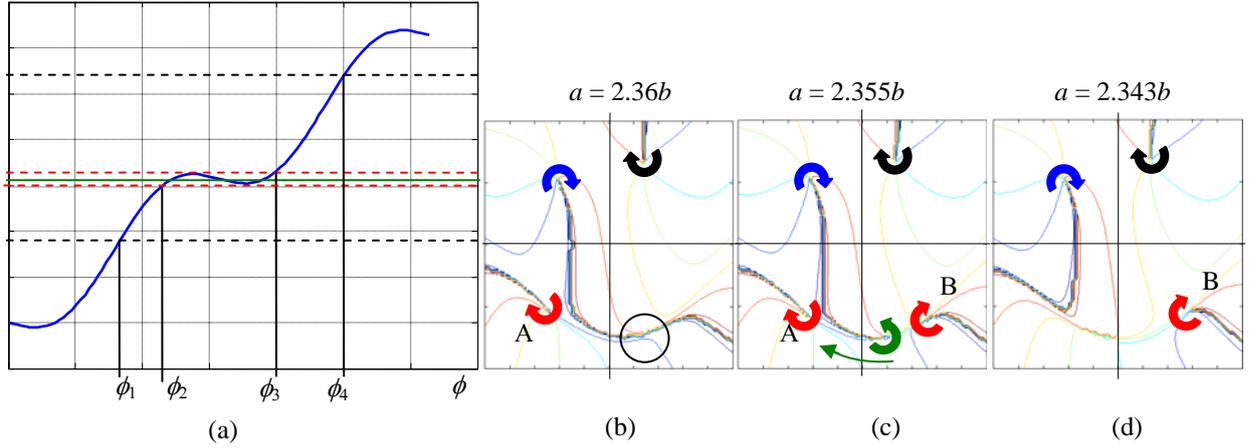

Fig. 3. (a) Illustration for solving the Eq. (4): The blue curve is the plot of the left-hand side expression at $|kra/(mz)| = 1.4$, horizontal lines symbolize different $a$-dependent values of the right-hand side. (b – d) Equiphase contours and the secondary OV positions in the diffracted beam cross section $z = 30$ cm for the Kummer beam with $b = 0.232$ mm [10,11] for the screen-edge positions indicated above (further explanations in text).

## IV. Discussion and conclusion

Therefore, the non-monotonic dependence of the left-hand side of Eq. (4) on $\phi$ completely explains the observed anomalies of the OV trajectories in the diffracted beam cross section. Obviously, this non-monotonic character may appear if

$$\left|\frac{kra}{mz}\right| > 1 \qquad (6)$$

which provides the criterion at which the jumps in the secondary OV trajectories within the diffracted beam cross section are possible. For large enough $a$, when the approximation of Eqs. (1) – (4) is valid and the OV displacement from the nominal beam axis is small ($r < b$), this criterion can be realized at not very high $z$. For conditions of [10,11] $k \approx 10^5$ cm, $b = 0.232$ mm, $z = 30$ cm, $a = 2.35b$, $r \approx 0.72b$ one finds $|kra/mz| \approx 1.0$, that is, the jump in the OV-core trajectory is a rather expected phenomenon. Also, criterion (6) explains why the numerical analysis reveals the jump anomalies at $z = 30$ cm but they cannot be detected at higher distances from the screen, e.g. at $z = 82$ cm.

Interesting to note, the process illustrated by Figs. 3b–d shows that the OV migration over the diffracted beam cross section can be realized not only by "physical" motion of an amplitude zero with sequentially passing all the intermediate points but also in a "virtual" manner, via generation of a dipole pair in a remote point and subsequent annihilation of the OV in the initial point. In such processes, the OV positions appear to be very sensitive to the smallest changes in the screen edge location, which potentially can be used in the optical-vortex "singularimetry" [13] for precise measurements of small displacements and deformations.